# UV-Vis-IR Spectral Complex Refractive Indices and Optical Properties of Brown Carbon Aerosol from Biomass Burning

Benjamin J. Sumlin[a]*, Yuli W. Heinson[a]*, Nishit Shetty[a], Apoorva Pandey[a], Robert S. Pattison[b], Stephen Baker[c], Wei Min Hao[c], and Rajan K. Chakrabarty[a]**

[a]Center for Aerosol Science and Engineering, Department of Energy, Environmental and Chemical Engineering, Washington University in St. Louis, St. Louis, MO 63130

[b]United States Forest Service, Pacific Northwest Research Station, Anchorage, AK 99501

[c]Missoula Fire Sciences Laboratory, United States Forest Service, Missoula, MT 59808

\* These authors contributed equally to this work.

\** Corresponding author: Rajan K. Chakrabarty (chakrabarty@wustl.edu)

# Abstract

Constraining the complex refractive indices, optical properties and size of brown carbon (BrC) aerosols is a vital endeavor for improving climate models and satellite retrieval algorithms. Smoldering wildfires are the largest source of primary BrC, and fuel parameters such as moisture content, source depth, geographic origin, and fuel packing density could influence the properties of the emitted aerosol. We measured *in situ* spectral (375-1047 nm) optical properties of BrC aerosols emitted from smoldering combustion of Boreal and Indonesian peatlands across a range of these fuel parameters. Inverse Lorenz-Mie algorithms used these optical measurements along with simultaneously measured particle size distributions to retrieve the aerosol complex refractive indices ($m=n+i\kappa$). Our results show that the real part *n* is constrained between 1.5 and 1.7 with no obvious functionality in wavelength ($\lambda$), moisture content, source depth, or geographic origin. With increasing $\lambda$ from 375 to 532 nm, $\kappa$ decreased from 0.014 to 0.003, with corresponding increase in single scattering albedo (SSA) from 0.93 to 0.99. The spectral variability of $\kappa$ follows the Kramers-Kronig dispersion relation for a damped harmonic oscillator. For $\lambda \geq 532$ nm, both $\kappa$ and SSA showed no spectral dependency. We discuss differences between this study and previous work. The imaginary part $\kappa$ was sensitive to changes in FPD, and we hypothesize mechanisms that might help explain this observation.

# 1. Introduction

Organic aerosols (OA) account for a large fraction of the total tropospheric particulate matter burden [1, 2]. These aerosols have been typically considered to predominantly scatter light in the visible solar spectrum. However, findings from field [3, 4] and laboratory studies [5, 6] show that a class of OA, optically defined as brown carbon (BrC), significantly absorb in the shorter visible wavelengths ($\lambda \sim$ 350-550 nm) with absorption Ångström exponents (AAE) ranging between 2 and 12 [7]. BrC aerosols have physical, chemical, and optical properties distinct from black carbon (BC) aerosols. BC has a fractal-like morphology with a deep black appearance caused by a significant, non-zero imaginary part $\kappa$ of its complex refractive index (RI) that is wavelength-independent over the visible and near-visible wavelengths [8]. In contrast, BrC aerosols are spherical in morphology and are yellow-brown in color due to values of $\kappa$ that increase sharply toward shorter visible and ultraviolet wavelengths. Constraining and parameterizing the spectral optical properties and RIs of BrC aerosols across the solar spectrum has been a challenging endeavor for the atmospheric aerosol community. Climate models and satellite retrieval algorithms rely on this information for accurate retrievals and predictions of aerosol optical depths.

Primary BrC aerosol emissions are largely attributed to biomass and biofuel burning [9-11] and biogenic release of soil and humic matter [7, 12]. In particular, it is the smoldering phase of biomass burning that has been identified as the major source of these particles [10, 13, 14]. Recent studies show that boreal and Indonesian peat fires are the largest contributors of primary BrC aerosols to regional emissions, and contribute up to 72% of all carbon emissions in a given year [15]. Peatlands store between one-fifth and one-third of earth's organic carbon and act as a net carbon sink, but this carbon is increasingly released back to the atmosphere since peatlands face an increasing threat of wildfires due to rising global temperatures [16-18]. Peat fires are dominated by smoldering phase combustion which can persist in low to moderate fuel moisture conditions, and is capable of lasting for several weeks or longer [19].

In this study, we present our results from *in situ*, contact-free measurements of spectral (UV-Vis-IR) optical properties and size distributions of BrC aerosol emitted from laboratory-scale smoldering combustion of peat samples collected from different parts of Alaska and Indonesia. Scattering and absorption coefficients $\beta_{sca}$ and $\beta_{abs}$ were measured using four integrated photoacoustic-nephelometers (IPNs). In conjunction with size distribution measurements by a scanning mobility particle sizer (SMPS), these optical measurements were inverted using Mie theory for the retrieval of complex RIs ($m=n+i\kappa$). Best efforts were made to mimic real-world smoldering fire scenarios in our laboratory experiments. Smoldering fire behavior fluctuates across a typical forest floor because of spatial variability in fuel depth, fuel packing density (FPD, mass per unit volume), mineral content, and moisture content (MC) [20-26]. The probability that peat will burn and sustain once ignited depends heavily on MC and FPD. We studied the variation in optical properties and size distributions of emitted smoke aerosols as a function of varying fuel depths, FPDs, and MCs. When compared to previous studies, we find that FPD has the strongest effect on BrC absorption properties, with $\kappa$ varying directly with FPD. Finally, we constrain previous literature results and our experimental findings on κ(λ) using the analytical form of the Kramers-Kronig dispersion relation for a damped harmonic oscillator [27].

## 2. Experimental methods

Fig. 1 shows the schematic diagram of our experimental setup. The setup consists of a sealed 21 m$^3$ stainless steel chamber equipped with a computer-controlled ignition system and a recirculation fan. The ignition system is a 1 kW ring heater (McMaster-Carr 2927094A) coupled to a 1/16" stainless steel plate, and its temperature is monitored by a K-type thermocouple to close the control loop. We studied peat samples collected from Alaska (AK) and Indonesia (IN). The AK peat samples were separated into collection depths of 0-4" and 4-8" below the surface from sites dominated by sphagnum and black spruce (*Picea mariana*). Typically, canopy cover of black spruce was about 40%. The understory was typically sparse, with species such as dwarf birch (*Betula nana*) and varieties of *Rhododendron* subsect. *Ledum*, *Vaccinium* and *Empetrum*. The AK peat samples were naturally dried to 5%, 10%, 15%, 20%, and 40% MC at room temperature. The IN peat samples were not depth-resolved and were dried to 5%, 20%, and 40% MCs. IN forest speciation information was unavailable due to the high degree of biodiversity in southeast Asian rainforests. Approximately 2 g of each fuel sample was placed on the heating plate such that the FPD was ~0.03 g/cm$^3$, and smoldering was initiated by heating the plate to 245 °C. One hour after ignition, aerosols were sampled from ports approximately 2 m above the chamber floor. Gas-phase products were removed with activated parallel-plate semivolatile organic carbon (SVOC) denuders, and excess water was removed with a diffusion dryer packed with indicating silica beads. Finally, the aerosols were mixed in a 208 liter barrel and a homogeneous, isokinetic stream was sampled by each IPN and the SMPS.

The IPNs used in this study are described in detail in the Supporting Material. We operated four custom designed IPNs at λ = 375, 405, 532, and 1047 nm. The IPNs measured $β_{sca}$ and $β_{abs}$ continuously with 2 sec time resolution, from which the single scattering albedo (SSA) was calculated. The SMPS measured particle number size distributions every 5-minutes. IPN measurements were averaged over 5 minutes to align with the SMPS measurement intervals.

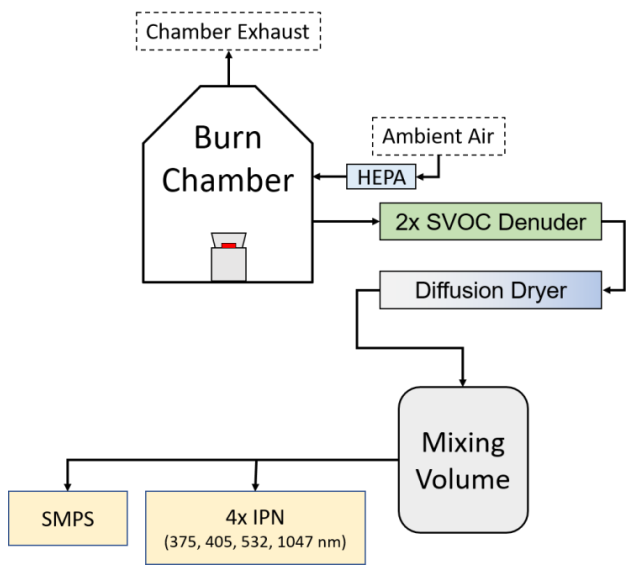

**Fig. 1**. A schematic diagram of the experimental setup showing the SVOC denuders, diffusion dryer, mixing volume, SMPS, and IPNs.

Complex RI was retrieved using PyMieScatt, a Lorenz-Mie theory package for Python 3 [28]. PyMieScatt includes inversion functions that take measurements of $\beta_{abs}$, $\beta_{sca}$, and the size distribution to return $m$. The theory behind the retrieval algorithms is detailed in Ref. 26. To minimize computing overhead, we chose PyMieScatt's Survey-Iteration algorithm. Briefly, this is a two-stage algorithm that first constructs coarse arrays of $\beta_{abs}(n,\kappa)$ and $\beta_{sca}(n,\kappa)$ for a given size distribution and wavelength of light, and surveys them for values that are close to the IPN measurements. Values with array indices that are common to both the $\beta_{abs}(n,\kappa)$ and $\beta_{sca}(n,\kappa)$ arrays are considered candidate solutions. The iteration stage is best described by Fig. 2. The real part of the refractive index is treated first in the red "Scattering" loop, and then the imaginary part is treated by the blue "Absorption" loop. The algorithm runs for each candidate $m$ found by the survey. When a solution is found, the algorithm reports $m$ along with $E_{abs}$ and $E_{sca}$, which are the residuals between simulated and measured $\beta_{abs}$ and $\beta_{sca}$.

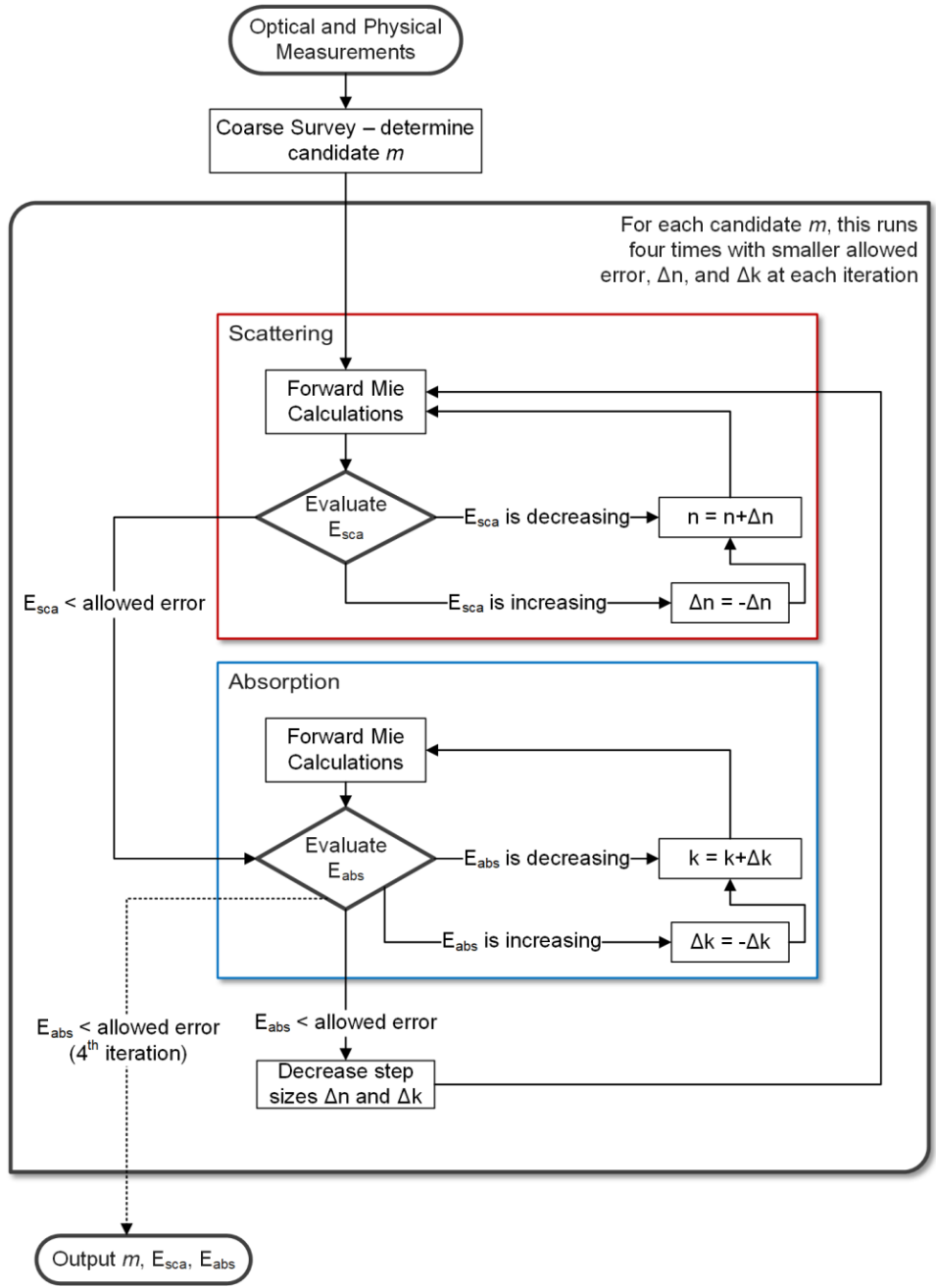

**Fig. 2.** Flowchart describing the iteration phase of the survey-iteration algorithm. $E_{abs}$ and $E_{sca}$ are the relative errors between measured and simulated $\beta_{abs}$ and $\beta_{sca}$.

We report $n$ and $\kappa$ as the average of twelve individual retrievals performed in 5-minute data intervals over an hour. Uncertainties for n, κ, SSA, AAE, and κ Ångström exponent (κAE) are propagated through the retrievals and we report the overall standard deviation using

$$\sigma_{\text{overall}} = \sqrt{\frac{\sum_i^N \sigma_i^2}{N}}, \qquad \text{(Eq. 1)}$$

where $\sigma_i$ is the standard deviation of each data point and N is the total number of data points.

## 3. Results and discussion

In this section, we refer to AK peat samples by their MCs and source depths. The IN samples are labelled according to their MCs. The data for all graphs is tabulated in the Supporting Material.

### 3.1 Complex Refractive Indices

The spectral dependence of *m* is plotted in Fig. 3. For both AK and IN peat, *n* is constrained between 1.5 and 1.7, and the spread of data belies any obvious trend in λ. However, we note that *n* is largely independent of MC, source depth, and geographic origin. We observed no functionality in *n* or κ as functions of anything other than λ, although in Section 4 we discuss differences in *m* under varying fuel packing densities.

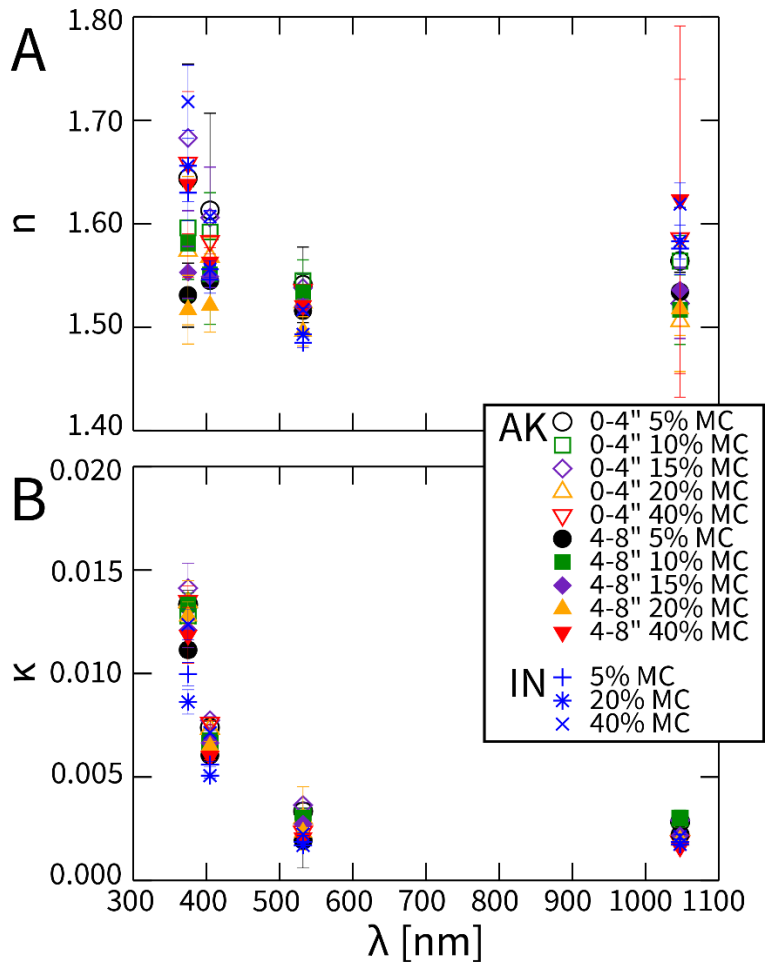

**Fig. 3.** A: Values of *n* show no significant dependence on λ, are constrained between 1.5 and 1.7, and are independent of the source depth, MC, and geographic origin. B: κ decreases monotonically with λ between 375 and 532 nm, above which it levels off at κ ≈ 0.002.

The current understanding of BrC optical properties has been summarized in previous work, notably by Liu et al. (2015) and Laskin et al. (2015) [7, 29]. Liu et al. compared the wavelength-dependent κ values for a variety of atmospheric light absorbing organic material such as *m*-xylene and toluene oxidation products [30, 31], as well as BrC from previous studies [32-38]. The κ values we report here, and their sensitivity to change in λ, are commensurate with those previously reported. In fact, when considering κ from this work and averages across the literature referenced by Liu et al., we find that κ(λ) follows the Kramers-Kronig dispersion relation (KK) for a damped harmonic oscillator [39]. Fig. 4 shows data from this work and the mean of values taken from literature, overlaid with the analytical form of KK given by:

$$\kappa = \frac{a\,\gamma\,\nu}{(\nu_0 - \nu)^2 + (\gamma\nu)^2} \qquad \text{(Eq. 2)}$$

where *a* is a constant, γ is a line width parameter, *ν* is the frequency of incident light, and $\nu_0$ is the resonance frequency of the oscillator. Fig. 4 uses $a = 10^{29}$ s$^{-2}$ and $\nu_0 = c/\lambda_0$, where $\lambda_0 = 300$ nm. The line width parameter γ was set to $2 \times 10^{13}$ s$^{-1}$. This function is appropriate near resonance for the BrC absorption spectrum at 375 ≤ λ ≤ 532 nm; far from resonance, i.e. λ = 1047 nm, average κ from literature and this work is 0.002±0.005.

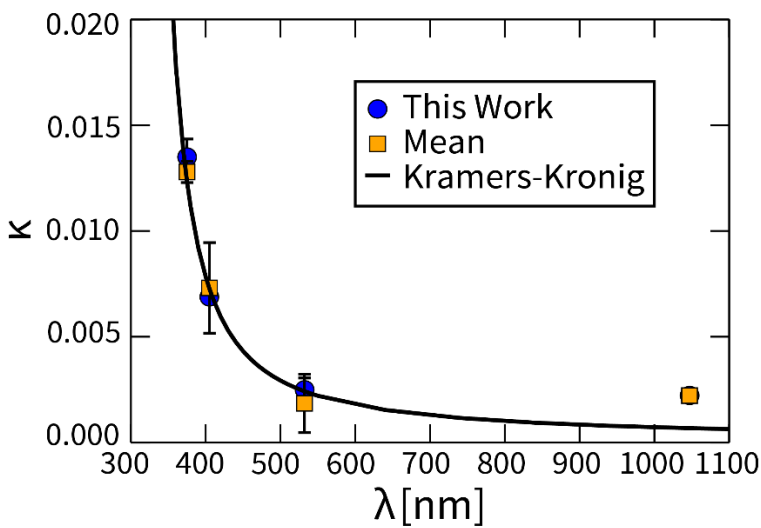

**Fig. 4.** κ values and the Kramers-Kronig dispersion relation per Eq. 2. The relationship is valid for λ up to 532 nm, beyond which κ appears constant.

### 3.2 Ångström Exponents

Ångström exponents (AE) can be used to describe the wavelength functionality of any optical parameter that follows a well-behaved power law [39]. For wavelength-independent parameters, the AE will be between zero and unity, and larger values indicate increasing sensitivity to changes in wavelength [40]. Fig. 5 demonstrates the wavelength dependence of BrC absorption by plotting the κ Ångström exponent (κAE) and absorption Ångström exponent (AAE) in three intervals of λ (375-405 nm, 405-532 nm, and 532-1047 nm) using the two-wavelength formula:

$$AE(\lambda_1, \lambda_2) = -\frac{\ln\left[\frac{P(\lambda_1)}{P(\lambda_2)}\right]}{\ln\left[\frac{\lambda_1}{\lambda_2}\right]} \quad \text{(Eq. 3)}$$

where $P(\lambda)$ is the wavelength-dependent parameter in question, either κ for κAE or $\beta_{abs}$ for AAE.

Previous studies have placed BrC κAE between 4 and 11 [39]. In this study, we find that BrC from peat smoldering is extremely sensitive to wavelength in the near-UV, with a 375-405 nm κAE values ranging between 7 and 9, while the AAE values range between 8 and 11. Both κAE and AAE decrease with increasing λ.

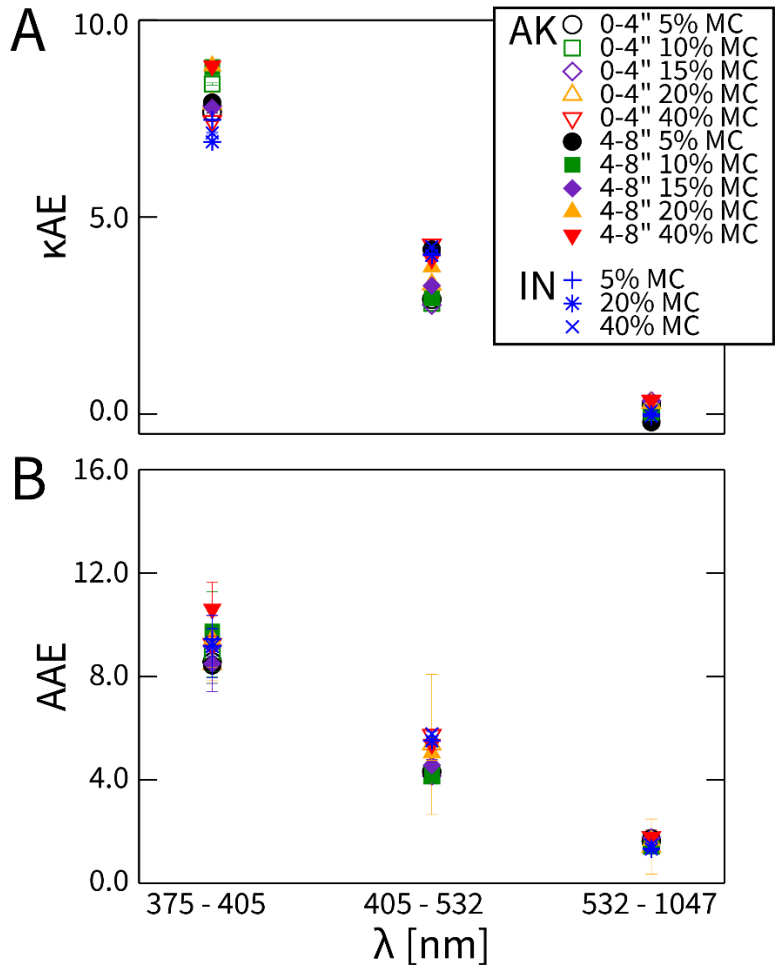

**Fig. 5.** Ångström exponents as a function of wavelength λ. **A:** κ Ångström exponent (κAE); **B:** absorption Ångström exponent (AAE).

### 3.3 Single Scatter Albedo

Given strong κ and weak *n* functionality with λ, SSA(λ) should follow an increasing trend with λ. Indeed, we observe such a trend in Fig. 6. SSA increases with λ until it reaches 0.99 for λ ≥ 532 nm for all peat samples.

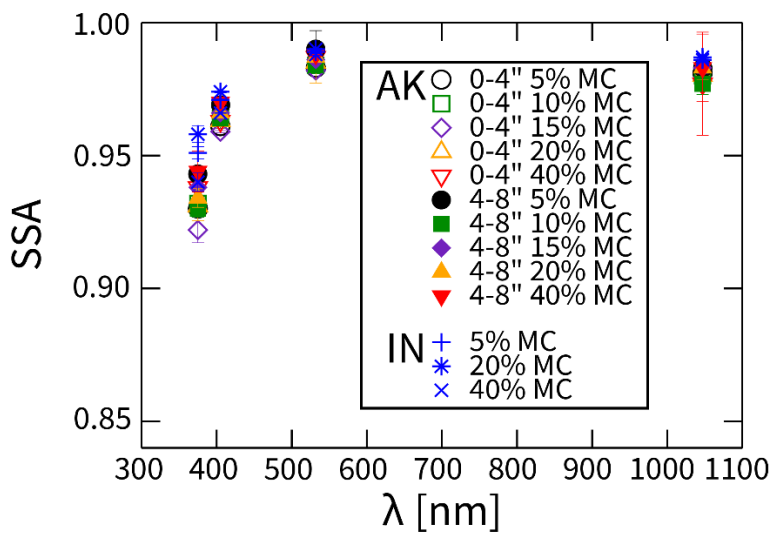

**Fig. 6.** SSA increases with λ until it reaches a value of 0.99 at λ ≥ 532 nm.

## 4. Comparisons to previous studies

We compared our findings against previous work that used nearly identical experimental methods to study the effects of atmospheric oxidation on the optical properties of BrC from smoldering combustion [41]. In Ref. [41], the control experiment (fresh, unoxidized BrC) was performed identically to the experiments in this study using AK peat at 5% MC and 0-4" source depth. In the present work, we have shown that BrC optical properties have no functional dependence on MC, source depth, or geographic origin. However, upon comparison to the work in Ref. [41], we find evidence that the fuel packing density (FPD) directly affects *m*. In this work, fuel was combusted at a packing density of ~0.03 g cm$^{-3}$, while in [41], the FPD was ~0.06 g cm$^{-3}$, which we term "Low FPD" and "High FPD", respectively. Fig. 7 compares the average *n*, κ, κAE and AAE of all experiments done in this study to the unaltered BrC emissions from Ref. [41], and in the case of κAE, literature results as well. On average, with higher FPD, *n* is slightly smaller while κ is significantly larger, which in turn increases both κAE and AAE. We find κ to be 3 times higher at 375 nm, and 1.5 times higher at 405 nm. Fig. 8 shows the comparisons of SSA, a key parameter for climate modelling and satellite retrievals. Densely packed peat, upon smoldering, emits BrC with a lower 375-405 nm SSA, while at λ ≥ 532, both low and high FPD peat BrC have SSA values reaching approximately 0.99.

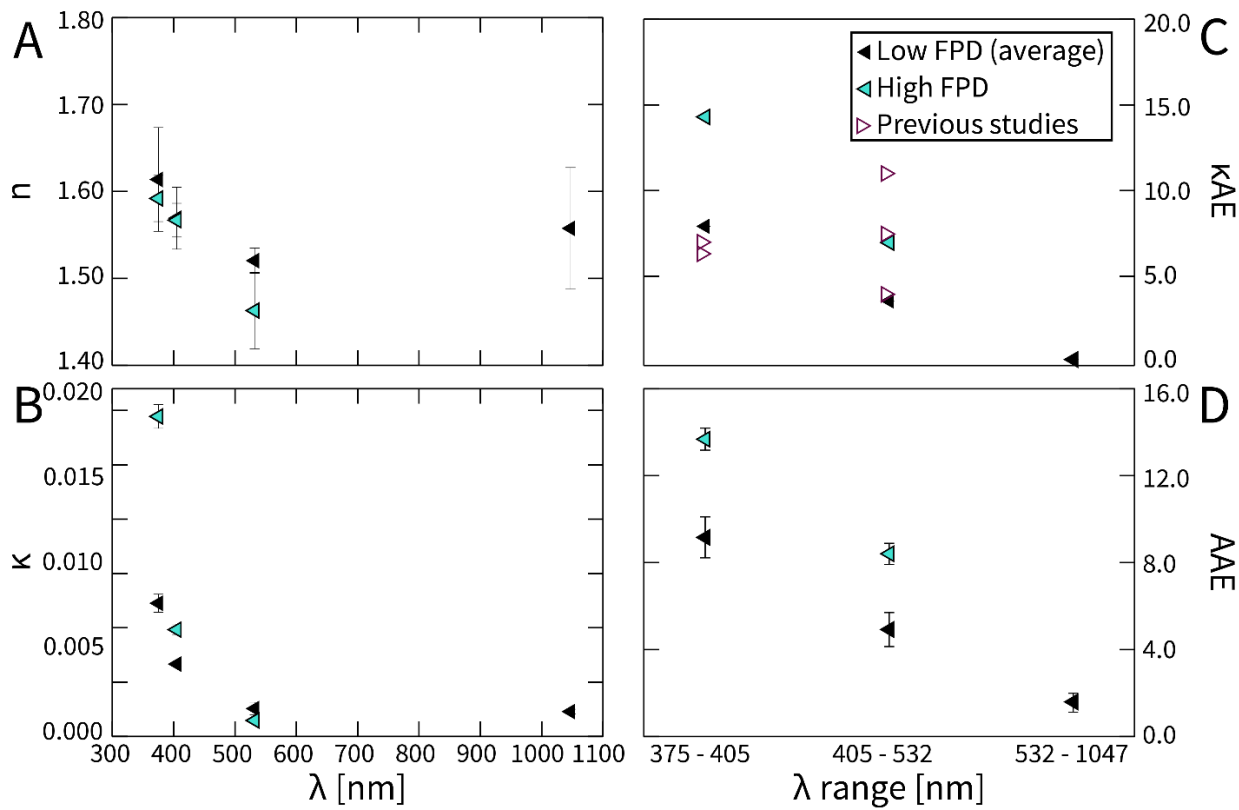

**Fig. 7.** Comparison of the average results from this study (low FPD, black triangles) to Ref. [41] (high FPD, blue triangles) and other literature results (magenta triangles). **A**: real refractive index *n*; **B**: imaginary refractive index κ; **C**: κAE; **D**: AAE. BrC from high FPD produced more absorbing and less scattering aerosol, indicated by smaller *n* and larger κ.

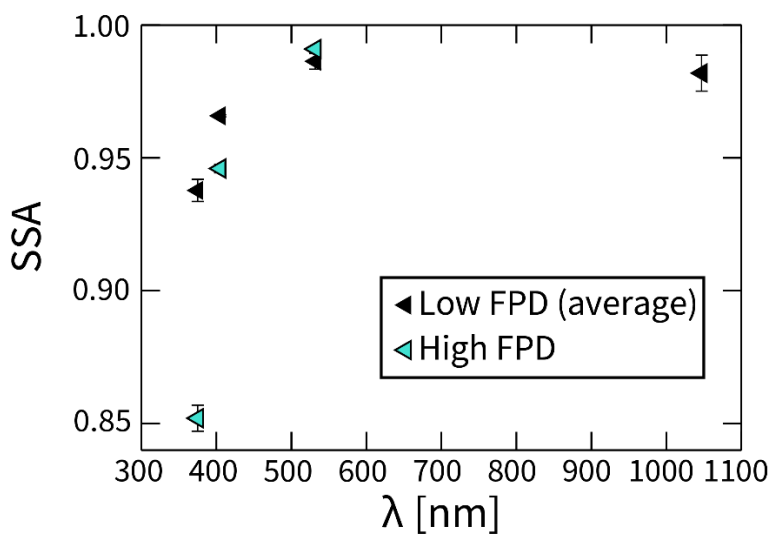

**Fig. 8.** Comparison of the average SSA from this study (low FPD, black triangles) to Ref. [41] (high FPD, blue triangles). As expected, SSA from both studies increases with wavelength,

although the larger near-UV κ from Ref. [41] results in a lower SSA. At λ ≥ 532, SSA from both studies is nearly 0.99.

Smoldering fire behavior fluctuates across a typical forest floor because of spatial variability in fuel depth, fuel packing density (FPD), mineral content, and MC [20, 22, 23, 42-44]. Combustion in a thick bed of peat is typically stratified, and flame front velocities depend on the depth at which the fire occurs [45]. We hypothesize that denser fuel packing would slow the velocity of the flame front through the fuel layers. The preheat zone ahead of the flame front would then expand as heat is transported away into uncombusted fuel. This larger preheat zone may volatilize organic compounds ahead of combustion, creating a VOC-rich zone entrained within the fuel through which particles created by combustion must travel, with the least volatile compounds emitted just ahead of combustion. These VOCs may condense on the particle, altering their mean molecular weight and speciation, and consequently modify their absorption behavior. However, FPD may simply be a proxy for oxygen availability. When the FPD was ~0.06 $g/cm^3$–corresponding to a 3/4" thick fuel packing on the heating plate–the smoldering front propagated slowly due to limited oxygen, which, by the mechanisms hypothesized, may produce higher molecular weight compounds with larger κ values. Conversely, when the FPD was ~0.03 $g/cm^3$, the environment was comparatively oxygen-rich, resulting in more rapid combustion and production of lower molecular weight compounds with smaller κ values. These hypotheses will be a subject of future research.

## 5. Concluding remarks and future work

We conducted *in situ* contact-free measurements of spectral optical properties of primary BrC aerosols emitted from smoldering combustion of peat samples collected from different parts of Alaska and Indonesia. Using optical and physical measurements, we retrieved the complex RIs of those aerosols using an inverse Mie algorithm and sought dependencies on MC, source depth, and geography. We found that BrC optical properties are not sensitive to these parameters, suggesting that climate models and satellite retrievals can make use of a smaller parameter space when considering BrC aerosol emitted from biomass burning. However, we found a novel relationship between the complex RIs and the FPD, with more densely packed fuel producing more highly-absorbing aerosols.

We expect FPD will vary widely in a real-world peat bog. The degree of fuel packing may vary across many orders of magnitude in a small area, and therefore future work is needed to draw quantitative conclusions between FPD and its effect on the optical properties of the emitted BrC, as well as to understand what this may mean for the broader impacts of BrC from smoldering wildfires.

## Acknowledgements

This work was partially supported by the National Science Foundation under Grant No. AGS1455215, NASA ROSES under Grant No. NNX15AI66G, and the International Center for Energy, Environment and Sustainability (InCEES) at Washington University in St. Louis.

## Data Availability

Tabulated data is available in the SM. The computer code used for refractive index retrievals is described in [28] and is available at https://github.com/bsumlin/PyMieScatt. Raw data is available from the authors upon request.

**Supporting Material for**

# UV-Vis-IR Spectral Complex Refractive Indices and Optical Properties of Brown Carbon Aerosol from Biomass Burning

Benjamin J. Sumlin[a]*, Yuli W. Heinson[a]*, Nishit Shetty[a], Apoorva Pandey[a], Robert S. Pattison[b], Stephen Baker[c], Wei Min Hao[c], and Rajan K. Chakrabarty[a]**

[a]Department of Energy, Environmental and Chemical Engineering, Washington University in St. Louis, St. Louis, MO 63130

[b]United States Forest Service, Pacific Northwest Research Station, Anchorage, AK 99501

[c]Missoula Fire Sciences Laboratory, United States Forest Service, Missoula, MT 59808

\* These authors contributed equally to this work.

\*\* Corresponding author: Rajan K. Chakrabarty (chakrabarty@wustl.edu)

# 1. Instrumentation

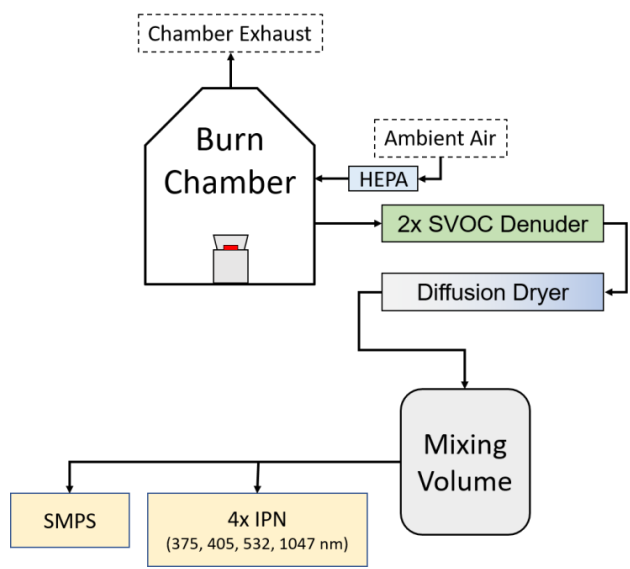

**Fig. S1.** The experimental setup.

Fig. S1 shows the schematic diagram of our experimental setup. The setup consists of a sealed 21 m$^3$ stainless steel chamber equipped with a computer-controlled ignition system and a recirculation fan. The ignition system is a 1 kW ring heater (McMaster-Carr 2927094A) coupled to a 1/16" stainless steel plate, and its temperature is monitored by a K-type thermocouple to close the control loop. One hour after ignition, aerosols were sampled from ports approximately 2 m above the chamber floor. Gas-phase products were removed with activated carbon parallel-plate denuders (Sunset Laboratory Inc., Tigard, OR), and excess water was removed with a diffusion dryer packed with indicating silica beads (McMaster-Carr part 2181K97). Finally, the aerosols were mixed in a 208 liter barrel (McMaster-Carr part 4392T47). Sampling took place directly from the mixing volume from ports connected to the circumference of the barrel at half the barrel height. Instrumentation was comprised of four integrated photoacoustic spectrometer/nephelometers (IPN) at 375, 405, 532, and 1047 nm, a scanning mobility particle sizer (SMPS, TSI, Inc., Shoreview, MN). The residence time distribution for the experimental setup was measured, and from this information we determined that one hour after ignition was the optimal time frame to begin data collection.

## 1.1 Integrated photoacoustic-nephelometer (IPN)

Photoacoustic spectroscopy and nephelometry have been widely applied to measure absorption and scattering of light by aerosols, and working principles and calibration methods are well documented in the literature [1-14]. The integrated photoacoustic-nephelometer (IPN) spectrometer is an in-situ, real-time, contact-free measurement that gives highly precise values of light scattering and absorption coefficients $\beta_{sca}$ and $\beta_{abs}$ at a single wavelength. In this study, single-pass IPN spectrometers of our own design were used.

The IPN measures $\beta_{abs}$ by the photoacoustic effect and $\beta_{sca}$ via an integrating nephelometer. The photoacoustic portion was designed and calibrated based on Arnott et al. (1999), with additional considerations for design and optimization from Arnott et al. (2006) [2, 4]. It utilizes a ½-wavelength plane-wave longitudinal resonator with the microphone and calibration speaker placed at pressure antinodes.

The integrated nephelometer was designed and calibrated based on Penaloza (1999) and Abu-Rahmah et al. (2006), and has a truncation angle of approximately 5° [9, 14]. The sensor is a silicon photodiode with a Teflon cosine lens.

Data were acquired at 2 s intervals and instrument zeros were obtained every 300 measurements. The zeroing process involves automatically switching from sample flow to HEPA-filtered flow via a mechanically-controlled valve actuated by the instrument software. The background measurement is averaged over 30 s to improve the signal-to-noise ratio for zeroing. Since the zeroing filter does not effectively remove gaseous species, excess ozone and $NO_X$ compounds are included in the instrument backgrounds and their interferences with absorption measurements are neglected during regular sampling.

The resonant frequency $f_0$ and quality factor Q are explicit functions of ambient conditions (namely temperature, pressure, and relative humidity) which are subject to fluctuations, even in a climate-controlled laboratory study. The IPN zeroing process measures $f_0$ and Q during zeroing by playing a chirp into the cell and determining the frequency where sound pressure is a maximum.

Calibrations are performed as commonly done in the literature, with non-absorbing aerosols (salt) to calibrate scattering, and absorbing aerosols (kerosene soot) to calibrate absorption. The slope of a linear regression of scattering versus extinction is the calibration factor for scattering, while the slope of absorption versus extinction-minus-scattering is used for absorption.

Raw data was post-processed to 5-minute averages, commensurate with the scan times of the SMPS. These 5-minute averages were used to determine the complex refractive indices ($m=n+i\kappa$) and single-scatter albedo (SSA).

## 2. Tabulated Data

**Table S1.** Complex refractive indices ($m=n+i\kappa$) and single scatter albedo (SSA) for AK peat. Values are averages ± one standard deviation.

| λ (nm) | Depth (inches) | MC (%) | n | κ | SSA |
|---|---|---|---|---|---|
| 375 | 0 - 4 | 5 | 1.644 ± 0.110 | 0.01334 ± 0.00068 | 0.930 ± 0.002 |
| | | 10 | 1.596 ± 0.050 | 0.01279 ± 0.00092 | 0.932 ± 0.005 |
| | | 15 | 1.683 ± 0.070 | 0.01413 ± 0.00120 | 0.922 ± 0.005 |
| | | 20 | 1.574 ± 0.072 | 0.01344 ± 0.00104 | 0.931 ± 0.006 |
| | | 40 | 1.659 ± 0.069 | 0.01351 ± 0.00073 | 0.938 ± 0.004 |
| | 4 - 8 | 5 | 1.531 ± 0.031 | 0.01114 ± 0.00035 | 0.943 ± 0.002 |
| | | 10 | 1.581 ± 0.055 | 0.01334 ± 0.00055 | 0.930 ± 0.003 |
| | | 15 | 1.553 ± 0.025 | 0.01210 ± 0.00085 | 0.938 ± 0.004 |
| | | 20 | 1.517 ± 0.033 | 0.01278 ± 0.00068 | 0.934 ± 0.004 |
| | | 40 | 1.638 ± 0.090 | 0.01185 ± 0.00135 | 0.944 ± 0.008 |
| 405 | 0 - 4 | 5 | 1.613 ± 0.094 | 0.00739 ± 0.00031 | 0.961 ± 0.001 |
| | | 10 | 1.592 ± 0.038 | 0.00671 ± 0.00016 | 0.964 ± 0.001 |
| | | 15 | 1.606 ± 0.049 | 0.00774 ± 0.00018 | 0.982 ± 0.001 |
| | | 20 | 1.568 ± 0.019 | 0.00733 ± 0.00001 | 0.963 ± 0.001 |
| | | 40 | 1.583 ± 0.006 | 0.00762 ± 0.00010 | 0.963 ± 0.001 |
| | 4 - 8 | 5 | 1.545 ± 0.004 | 0.00606 ± 0.00012 | 0.969 ± 0.001 |
| | | 10 | 1.551 ± 0.048 | 0.00677 ± 0.00020 | 0.964 ± 0.001 |
| | | 15 | 1.549 ± 0.011 | 0.00665 ± 0.00001 | 0.966 ± 0.001 |
| | | 20 | 1.521 ± 0.026 | 0.00647 ± 0.00024 | 0.966 ± 0.001 |
| | | 40 | 1.563 ± 0.008 | 0.00600 ± 0.00016 | 0.970 ± 0.001 |
| 532 | 0 - 4 | 5 | 1.541 ± 0.036 | 0.00334 ± 0.00012 | 0.983 ± 0.001 |
| | | 10 | 1.545 ± 0.020 | 0.00312 ± 0.00023 | 0.984 ± 0.001 |
| | | 15 | 1.538 ± 0.013 | 0.00364 ± 0.00015 | 0.982 ± 0.001 |
| | | 20 | 1.524 ± 0.014 | 0.00299 ± 0.00153 | 0.985 ± 0.008 |
| | | 40 | 1.536 ± 0.004 | 0.00236 ± 0.00011 | 0.987 ± 0.001 |
| | 4 - 8 | 5 | 1.516 ± 0.003 | 0.00194 ± 0.00133 | 0.990 ± 0.007 |
| | | 10 | 1.534 ± 0.017 | 0.00304 ± 0.00022 | 0.984 ± 0.001 |
| | | 15 | 1.520 ± 0.007 | 0.00273 ± 0.00018 | 0.986 ± 0.001 |
| | | 20 | 1.496 ± 0.014 | 0.00233 ± 0.00023 | 0.987 ± 0.001 |
| | | 40 | 1.521 ± 0.003 | 0.00205 ± 0.00010 | 0.988 ± 0.001 |
| 1047 | 0 - 4 | 5 | 1.564 ± 0.052 | 0.00284 ± 0.00027 | 0.981 ± 0.002 |
| | | 10 | 1.564 ± 0.025 | 0.00301 ± 0.00023 | 0.980 ± 0.001 |
| | | 15 | 1.523 ± 0.034 | 0.00290 ± 0.00019 | 0.979 ± 0.002 |
| | | 20 | 1.506 ± 0.014 | 0.00247 ± 0.00022 | 0.980 ± 0.002 |
| | | 40 | 1.586 ± 0.154 | 0.00188 ± 0.00023 | 0.977 ± 0.019 |
| | 4 - 8 | 5 | 1.534 ± 0.019 | 0.00222 ± 0.00024 | 0.983 ± 0.002 |
| | | 10 | 1.517 ± 0.034 | 0.00299 ± 0.00042 | 0.977 ± 0.004 |
| | | 15 | 1.536 ± 0.022 | 0.00218 ± 0.00023 | 0.986 ± 0.001 |
| | | 20 | 1.518 ± 0.061 | 0.00194 ± 0.00015 | 0.983 ± 0.005 |

| | | 40 | 1.623 ± 0.168 | 0.00161 ± 0.00017 | 0.983 ± 0.013 |

**Table S2.** *m* and SSA for IN peat. Values are averages ± one standard deviation.

| λ (nm) | MC (%) | n | κ | SSA |
|---|---|---|---|---|
| 375 | 5 | 1.630 ± 0.027 | 0.00997 ± 0.00058 | 0.951 ± 0.002 |
| | 20 | 1.656 ± 0.034 | 0.00863 ± 0.00059 | 0.958 ± 0.003 |
| | 40 | 1.718 ± 0.035 | 0.01237 ± 0.00075 | 0.940 ± 0.001 |
| 405 | 5 | 1.546 ± 0.013 | 0.00561 ± 0.00013 | 0.971 ± 0.001 |
| | 20 | 1.556 ± 0.008 | 0.00507 ± 0.00012 | 0.974 ± 0.001 |
| | 40 | 1.607 ± 0.006 | 0.00714 ± 0.00010 | 0.966 ± 0.001 |
| 532 | 5 | 1.485 ± 0.005 | 0.00186 ± 0.00001 | 0.989 ± 0.001 |
| | 20 | 1.493 ± 0.002 | 0.00169 ± 0.00001 | 0.990 ± 0.001 |
| | 40 | 1.517 ± 0.002 | 0.00224 ± 0.00001 | 0.988 ± 0.001 |
| 1047 | 5 | 1.576 ± 0.010 | 0.00186 ± 0.00001 | 0.986 ± 0.001 |
| | 20 | 1.583 ± 0.032 | 0.00175 ± 0.00001 | 0.987 ± 0.001 |
| | 40 | 1.619 ± 0.0205 | 0.00212 ± 0.00011 | 0.986 ± 0.001 |

**Table S3.** Absorption Ångström Exponents (AAE) for AK peat. Values are averages ± one standard deviation.

| λ range (nm) | Depth (inches) | MC (%) | AAE |
|---|---|---|---|
| 375 - 405 | 0 - 4 | 5 | 8.566 ± 0.404 |
| | | 10 | 8.941 ± 0.949 |
| | | 15 | 9.188 ± 0.808 |
| | | 20 | 8.483 ± 0.621 |
| | | 40 | 9.256 ± 1.093 |
| | 4 - 8 | 5 | 8.425 ± 0.689 |
| | | 10 | 9.740 ± 1.541 |
| | | 15 | 8.494 ± 1.079 |
| | | 20 | 9.448 ± 1.086 |
| | | 40 | 10.600 ± 1.040 |
| 405 – 532 | 0 - 4 | 5 | 4.326 ± 0.172 |
| | | 10 | 4.155 ± 0.290 |
| | | 15 | 4.154 ± 0.188 |
| | | 20 | 5.371 ± 2.710 |
| | | 40 | 5.737 ± 0.226 |
| | 4 - 8 | 5 | 4.275 ± 0.263 |
| | | 10 | 4.181 ± 0.314 |
| | | 15 | 4.566 ± 0.229 |
| | | 20 | 5.043 ± 0.290 |
| | | 40 | 5.382 ± 0.209 |
| 532 - 1047 | 0 - 4 | 5 | 1.641 ± 0.125 |
| | | 10 | 1.450 ± 0.197 |
| | | 15 | 1.755 ± 0.127 |

| | | 20 | 1.421 ± 1.067 |
| | | 40 | 1.796 ± 0.266 |
| | | 5 | 1.740 ± 0.164 |
| | | 10 | 1.467 ± 0.267 |
| | 4 - 8 | 15 | 1.758 ± 0.210 |
| | | 20 | 1.659 ± 0.204 |
| | | 40 | 1.773 ± 0.131 |

**Table S4.** AAE for IN peat. Values are averages ± one standard deviation.

| λ range (nm) | MC (%) | AAE |
|---|---|---|
| 375 - 405 | 5 | 9.439 ± 0.436 |
| | 20 | 9.165 ± 1.208 |
| | 40 | 9.305 ± 0.519 |
| 405 - 532 | 5 | 5.503 ± 0.087 |
| | 20 | 5.510 ± 0.148 |
| | 40 | 5.781 ± 0.075 |
| 532 - 1047 | 5 | 1.355 ± 0.122 |
| | 20 | 1.331 ± 0.171 |
| | 40 | 1.473 ± 0.128 |

**Table S5.** κ Ångström Exponents (κAE) for AK peat. Values are averages ± one standard deviation.

| λ range (nm) | Depth (inches) | MC (%) | κAE |
|---|---|---|---|
| 375 - 405 | 0 - 4 | 5 | 7.675 ± 0.002 |
| | | 10 | 8.382 ± 0.002 |
| | | 15 | 7.821 ± 0.003 |
| | | 20 | 7.877 ± 0.002 |
| | | 40 | 7.441 ± 0.001 |
| | 4 - 8 | 5 | 7.911 ± 0.001 |
| | | 10 | 8.813 ± 0.001 |
| | | 15 | 7.778 ± 0.002 |
| | | 20 | 8.845 ± 0.002 |
| | | 40 | 8.843 ± 0.003 |
| 405 – 532 | 0 - 4 | 5 | 2.912 ± 0.001 |
| | | 10 | 2.808 ± 0.001 |
| | | 15 | 2.766 ± 0.001 |
| | | 20 | 3.228 ± 0.004 |
| | | 40 | 4.297 ± 0.001 |
| | 4 - 8 | 5 | 4.176 ± 0.005 |
| | | 10 | 2.935 ± 0.001 |
| | | 15 | 3.264 ± 0.001 |
| | | 20 | 3.744 ± 0.001 |
| | | 40 | 3.937 ± 0.001 |
| 532 - 1047 | 0 - 4 | 5 | 0.240 ± 0.000 |
| | | 10 | 0.053 ± 0.000 |

|  |  | 15 | 0.336 ± 0.000 |
|  |  | 20 | 0.282 ± 0.002 |
|  |  | 40 | 0.336 ± 0.000 |
|  | 4 - 8 | 5 | -0.199 ± 0.001 |
|  |  | 10 | 0.024 ± 0.001 |
|  |  | 15 | 0.332 ± 0.001 |
|  |  | 20 | 0.271 ± 0.000 |
|  |  | 40 | 0.357 ± 0.000 |

**Table S6.** κAE for IN peat. Values are averages ± one standard deviation.

| λ range (nm) | MC (%) | κAE |
| --- | --- | --- |
| 375 - 405 | 5 | 7.472 ± 0.001 |
|  | 20 | 6.911 ± 0.001 |
|  | 40 | 7.141 ± 0.001 |
| 405 - 532 | 5 | 4.047 ± 0.001 |
|  | 20 | 4.028 ± 0.001 |
|  | 40 | 4.250 ± 0.001 |
| 532 - 1047 | 5 | 0.000 ± 0.000 |
|  | 20 | -0.052 ± 0.000 |
|  | 40 | 0.081 ± 0.000 |

**Table S7.** Complex refractive index ($m=n+i\kappa$) and single scattering albedo (SSA) for high FPD AK peat (fuel packing density ~0.06 g cm$^{-3}$) from Sumlin et al. (2017) [15]. Values are averages ± one standard deviation.

| λ (nm) | n | κ | SSA |
| --- | --- | --- | --- |
| 375 | 1.592 ± 0.027 | 0.02945 ± 0.00110 | 0.852 ± 0.005 |
| 405 | 1.567 ± 0.019 | 0.00983 ± 0.00044 | 0.946 ± 0.002 |
| 532 | 1.463 ± 0.044 | 0.00147 ± 0.00016 | 0.991 ± 0.001 |

**Table S8.** Absorption Ångström Exponents (AAE) for high FPD AK peat from Sumlin et al. (2017). Values are averages ± one standard deviation.

| λ range (nm) | AAE |
| --- | --- |
| 375 - 405 | 13.678 ± 0.505 |
| 405 - 532 | 8.406 ± 0.484 |

**Table S9.** κ Ångström Exponents (κAE) for high FPD AK peat from Sumlin et al. (2017). Values are averages ± one standard deviation.

| λ range (nm) | κAE |
| --- | --- |
| 375 - 405 | 14.257 ± 0.005 |
| 405 - 532 | 6.967 ± 0.004 |